\documentstyle[psfig,prb,aps]{revtex} 
\begin{document}
\title{The physical origin of the electron-phonon
vertex correction} 
\author{C. Grimaldi$^{1,2}$, L. Pietronero$^{1,2}$ and M. Scattoni$^{1}$} 
\address{$^{1}$Dipartimento di Fisica, Universit\'{a} di Roma I ``La
Sapienza", 
Piazzale A.  Moro, 2, 00185 Roma, Italy }
\address{$^{2}$ Istituto Nazionale Fisica della Materia, Unit\'a di Roma 1, Italy}
\date{\today} 
\maketitle 
\begin{abstract}

The electron-phonon vertex correction has a complex 
structure both in momentum and frequency. We explain this structure 
on the basis of physical considerations
and we show how the vertex correction can be decomposed
into two terms with different physical origins.
In particular, the first term describes the
lattice polarization induced by the electrons and it is essentially
a single-electron process whereas the second term
is governed by the particle-hole excitations due to
the exchange part of the phonon-mediated electron-electron 
interaction. We show that by weakening the influence of
the exchange interaction the vertex takes mostly positive 
values giving rise to an enhanced effective coupling in the  
scattering with phonons. This weakening of the exchange 
interaction can be obtained by lowering the density of the
electrons, or by considering only long-ranged (small $q$) 
electron-phonon couplings.
These findings permit to understand why in the High-$T_c$ 
materials the small carrier density and the long ranged 
electron-phonon interaction may play a positive role in enhancing 
$T_c$.  
\\
\end{abstract}
\vskip 2pc  


\section{Introduction}
\label{intro}

In conventional metals, according to Migdal's theorem,\cite{migdal} 
the smallness of the parameter $\lambda\omega_D/E_F$ where
$\lambda$ is the electron-phonon coupling, 
$\omega_D$ and $E_F$ are typical phonon and electron energies respectively,
permits to describe succesfully the electron-phonon coupled system by 
neglecting the vertex corrections in the electronic self-energy.
The application of Migdal's theorem to the superconducting state has
led to the Migdal-Eliashberg (ME) theory of superconductivity, which
accurately describes the conventional superconductors.
However, in recent years, the discovery of complex materials showing
high-$T_c$ superconductivity, colossal magnetoresistance etc.
has raised doubts concerning the validity of Migdal's theorem
for such class of materials.
In fact, the fullerene compounds show  vibrational 
spectra ranging from few meV to about 0.2 eV, while the 
electronic conduction band has a width of approximately 0.5 eV.\cite{gunna1}
In this situation therefore the adiabatic parameter can be as large as
$\omega_D/E_F\simeq 0.8$.
Also for the cuprates the situation points toward the breakdown
of Migdal's theorem. For example, BSCCO compounds have
$\omega_D/E_F\simeq 0.2$ which is a small but not negligible 
value.\cite{uemura}
For these systems therefore the electronic and phononic 
dynamics have comparable energy scales and there is not an
{\it a priori} justification to neglect the vertex corrections.
\cite{pietro1,pietro2}

Recently, we have proposed a generalization of the ME theory
to include the first nonadiabatic effects.\cite{grima1,pietro3}
Our aim was to provide a scenario for the high-$T_c$ superconductivity
different from the ones arising from purely electronic or magnetic
pairing mechanisms. We were motivated by several clues such as
the absence of magnetic
ordering in doped fullerenes and the presence of large isotope effects in
underdoped cuprates.\cite{franck}
Moreover, the recent observation of an ion-mass dependence of
the penetration length in underdoped YBa$_2$Cu$_3$O$_{6+x}$ and
La$_{2-x}$Sr$_x$CuO$_4$ \cite{zhao} has provided a strong evidence
for the breakdown of Migdal's theorem.\cite{grima2}

A theoretical framework which goes beyond Migdal's limit
can lead to different situations.
For example, by enhancing the coupling to the lattice
vibrations one could end to polaron and eventually to
bi-polaron formation with the appearance of huge effective masses
for the charge carriers. Such a regime is beyond Migdal's limit,
however it unlikely gives rise
to high temperature superconductivity
and the observation of a Fermi
surface contraddicts the hypothesis of condensation of preformed bi-polarons
in its simplest form.
Instead our approach is a perturbative one, where the small parameter
is $\lambda\omega_D/E_F$. In this way the electron-phonon coupled system
is away from the polaron formation which is definitely a non-perturbative
approach.\cite{capone} One advantage of our method is to have 
an appreciable enhancement of the effective coupling in the Cooper
channel and at the same time reasonable values of charge carrier masses.
Moreover, the detailed study of the first nonadiabatic corrections to the
ME theory has unlighted the importance of the interplay between the
exchanged frequency $\omega$ and the momentum transfer $q$ in the 
electron-phonon scattering process. The first vertex correction is in fact 
very sensitive to the ratio $v_F q/\omega$ in such a way that
as long as $v_F q/\omega<1$ the vertex assumes positive values which
reflects to an enhancement of the critical temperature $T_c$.\cite{grima1,pietro3}
This strong $(\omega,q)$-dependence of the nonadiabatic corrections
leads also to interesting effects in the pressure coefficient of the
superconducting transition temperature\cite{sarkar} and in the
tunneling current in the superconducting state.\cite{umma}

Another striking feature is certainly the different effective
el-ph couplings associated to the normal state and the Cooper channel.
In fact the nonadiabatic theory of superconductivity introduces
two effective couplings, $\lambda_{\Delta}$ and $\lambda_Z$, which
in the adiabatic limit reduce both to $\lambda$, {\it i.e.}, the electron-phonon
coupling of the ME theory.
Such a result opens the possibility of having different effective
electron-phonon couplings associated to different physical quantities.
One example of such a peculiarity is given by the different dependence
of the coefficients of the isotope effects on $T_c$ and on
the effective charge carriers mass $m^*$
upon the adiabatic parameter $\omega_D/E_F$.\cite{grima2}

The features listed above are all consequences of the complex behavior
of the first vertex corrections (in the normal and superconducting states).
Therefore a detailed analysis of the nonadiabaticity in the electron-phonon
problem requires a detailed study of the nonadiabatic diagrams.
Other attempts in similar directions have been made also since before
the discovery of high-$T_c$ materials.
For example in Ref.\onlinecite{grabo}, the Eliashberg equations have been
modified by taking into account the first vertex corrections in the
context of plasmon-mediated superconductivity. However in this case
the vertex was considered only in the $\omega/v_F q<1$ regime 
which gives negative nonadiabatic contributions and lowers $T_c$.
Other authors have studied the vertex corrections using different
approximation schemes which apply
only in some particular region of the
($\omega,q$) plane.\cite{krishna,kostur,free,mierze}
Finally, calculations based on the Ward identity have led to
contrasting results. In Ref.\onlinecite{cai}, for example, a strong 
suppression of $T_c$ contrasts with the strong enhancement found 
in Ref.\onlinecite{takada}. 
All these contraddictory results come from particular
approximations which predilige different limits of the vertex function.
To be more specific, 
in Refs.\onlinecite{grabo,krishna,kostur,free,mierze,cai}
the $v_F q/\omega>1$ regime was considered while in
Ref.\onlinecite{takada} the vertex was evaluated
in the $v_F q/\omega<1$ limit.
So far, the most accurate results on vertex corrections are given by
numerical calculations which do not privilege any particular region
of the ($\omega,q$) plane.\cite{gunna2,perali}

It is certainly true that the value of  $v_F q/\omega$ depends on the characteristic
of the specific material one wants to investigate, however it is not completely
clear the reason of such a complex behavior of the vertex function. 
In this paper we try clarify this point by investigating the physics
behind the vertex function. Concerning the theory
of nonadiabatic superconductivity, the interpretation of the nonadiabatic
corrections in terms of physical processes will help to identify the characteristic
of the materials which can lead to an enhancement of the critical temperature.

In this paper we first summarize the behavior of the vertex function for
different values of the ratio $\omega_0/E_F$ and of the 
electron-density $n$.
In section \ref{limiting} we focus on the one-electron case and  
the anti-adiabatic limit $\omega_0\rightarrow\infty$. 
For these two cases the interpretation in terms
of physical mechanisms turns out to be particularly 
straightforward.
The last section is devoted to a general discussion
and conclusions.

\section{Behavior of the vertex function}
\label{behav}

In this section we consider the electron-phonon vertex correction
and its behavior
as a function of the adiabatic parameter $\omega_0/E_F$ and
the electron density $n$.
In the following analysis, we neglect the Coulomb interaction
between electrons. Therefore we
consider an Hamiltonian describing electrons 
with dispersion $\epsilon_{{\bf k}}$ interacting with phonons 
via a momentum dependent electron-phonon
matrix element $\gamma_{{\bf q}}$:

\begin{equation}
\label{Ham1}
H=\sum_{{\bf k},\sigma}\epsilon_{{\bf k}}
c^{\dagger}_{{\bf k}\sigma}c_{{\bf k}\sigma}+
\omega_0\sum_{{\bf q}}b^{\dagger}_{{\bf q}}b_{{\bf q}}+ \\
\sum_{{\bf k},{\bf q},\sigma}\gamma_{{\bf q}}c^{\dagger}_{{\bf k}\sigma}
c_{{\bf k}-{\bf q}\sigma}
\left(b_{{\bf q}}+b^{\dagger}_{-{\bf q}}\right).
\end{equation}
Here, $\omega_0$ is the phonon frequency, assumed to be
dispersionless for simplicity, and 
$c^{\dagger}_{{\bf k}\sigma}$ ($c_{{\bf k}\sigma}$) is the 
creation (annihilation) operator for an electron with wave number
vector ${\bf k}$ and spin $\sigma$ and $b^{\dagger}_{{\bf q}}$ 
($b_{{\bf q}}$) is the creation
(annihilation) operator for phonons with momentum ${\bf q}$.

The thermal Green's function for the electron $G(k)$ 
satisfies the usual Dyson equation:

\begin{equation}
\label{dyson}
G^{-1}(k)=G_0^{-1}(k)-\Sigma(k),
\end{equation}
where we have used the four-momentum representation $k=({\bf k},i\omega_k)$.
$\omega_k=(2n_k+1)\pi T$ is the fermionic Matsubara frequency and $T$ is
the temperature. In Eq.(\ref{dyson}), 
$G_0^{-1}(k)=i\omega_k-\epsilon_{{\bf k}}$ and $\Sigma(k)$ is
the electron self-energy due to the electron-phonon coupling:

\begin{equation}
\label{self1}
\Sigma(k)=\sum_q g^2_{{\bf q}}\Gamma(k+q,k)
D(q)G(k+q),
\end{equation}
where $g^2_{{\bf q}}=2\gamma^2_{{\bf q}}/\omega_0$ and

\begin{equation}
\label{di}
D(q)=D(i\omega_q)=\frac{\omega_0^2}{(i\omega_q)^2-\omega_0^2}
\end{equation}
is the phononic propagator with Matsubara frequencies $\omega_q=2n_q\pi T$.
In writing Eq.(\ref{self1}), we have used the short notation
$\sum_q=-(T/N)\sum_{\omega_q}\sum_{{\bf q}}$.
We consider the phonons as already renormalized and we
interpret Eq.(\ref{di}) in a phenomenological way.

\begin{figure}
\protect
\centerline{\psfig{figure=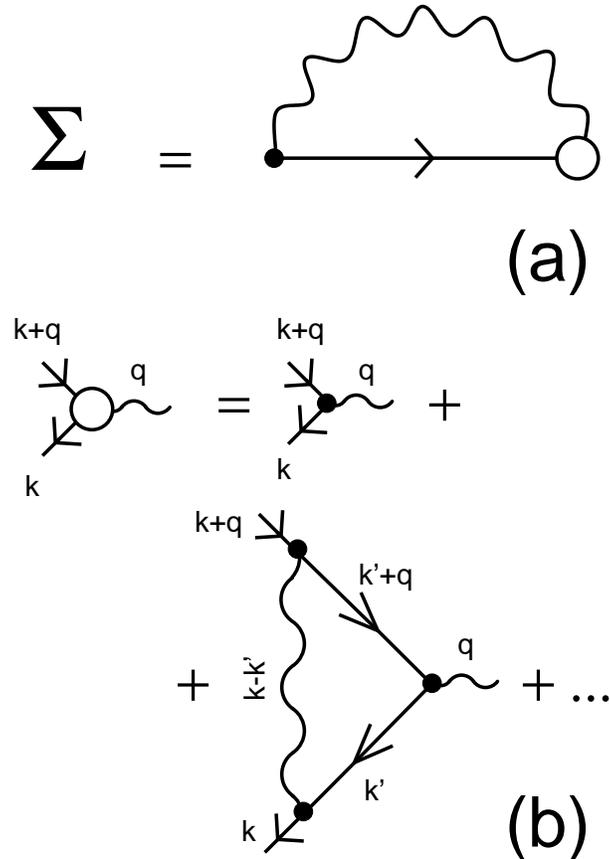,width=8cm}}
\caption{(a): electron self-energy. The open circle
represents the set $\Gamma$ of all irreducible vertex diagrams.
(b): expansion of $\Gamma$. The first diagram represents the bare
electron-phonon interaction $g({\bf q})$ while the second one
is the first vertex correction $g({\bf q})P(k+q,k)$ which in the
adiabatic limit gives a negligible contribution to $\Sigma$.} 
\label{fig1}
\end{figure}

In terms of Feynmann diagrams the self-energy (\ref{self1}) is shown in
Fig. \ref{fig1}a, where the solid  and wiggled lines are electronic and phononic
propagators, respectively. In Fig. \ref{fig1}a the open circle is the proper
vertex function
$\Gamma(k+q,k)$, which is 
given by all diagrams which cannot be separated into two different contributions
by cutting a single electron or phonon propagator line.
In Fig. \ref{fig1}b, we show the expansion of the vertex function up to the first
correction: $\Gamma(k+q,k)=1+P(k+q,k)$, where the vertex
correction $P(k+q,k)$ is given by
the following expression:

\begin{equation}
\label{vertice0}
P(k+q,k)=
\sum_{k'}g^2_{{\bf k}-{\bf k}'}
D(k-k')G(k'+q)G(k').
\end{equation}

The aim of the present paper is to provide a physical interpretation
of the above vertex correction. In this way, we should be able also to
interpret in terms of physical processes its complex behavior 
already pointed out in 
Refs.\onlinecite{pietro2,grima1,pietro3,capp1,capp2}
and that we remind here briefly.
First,
Migdal's theorem states that the order of magnitude of the vertex 
correction
(\ref{vertice0})
is  $P(k+q,k)\sim \lambda\omega_0/(v_Fq)$, where 
$\lambda\simeq \langle g^2_{{\bf q}}\rangle/E_F$
is the electron-phonon coupling and $v_F$ is the Fermi velocity.
Since in conventional metals $\lambda < 1$ and
the momentum transfer $q$ is of order of the Debye momentum
$q_D$, 
$\omega_0/(v_F q_D)\simeq \omega_0/E_F \ll 1$. The 
vertex correction $P$
can be therefore safely neglected and $\Gamma\simeq 1$.

However, when $\omega_0$ is of the same order of $E_F$ or when $q\ll q_D$
the vertex correction is no longer negligible. This situation can be
outlined by approximating the electron propagators $G$ with their free electron
form $G_0$. In this way, the sum over Matsubara frequencies in Eq.(\ref{vertice0})
can be performed exactly and the vertex correction reduces to the following form:

\begin{eqnarray}
\label{vertice1}
& & P({\bf k}+{\bf q},{\bf k};i\omega_k+i\omega_q, i\omega_k) = 
\frac{\omega_0}{2}\sum_{{\bf k}'}\frac{g^{2}_{{\bf k}-{\bf k}'}}
{\epsilon_{{\bf k}'}-\epsilon_{{\bf k}'+{\bf q}}+i\omega_q} \nonumber \\
& & \times\left[ \frac{f(\epsilon_{{\bf k}'})+n(-\omega_0)}
{\epsilon_{{\bf k}'}+\omega_0-i\omega_k}
-\frac{f(\epsilon_{{\bf k}'})+n(\omega_0)}
{\epsilon_{{\bf k}'}-\omega_0-i\omega_k} 
- \frac{f(\epsilon_{{\bf k}'+{\bf q}})+n(-\omega_0)}
{\epsilon_{{\bf k}'+{\bf q}}+\omega_0-i(\omega_k+\omega_q)}+
\frac{f(\epsilon_{{\bf k}'+{\bf q}})+n(\omega_0)}
{\epsilon_{{\bf k}'+{\bf q}}-\omega_0-i(\omega_k+\omega_q)}
\right] . \nonumber \\
\end{eqnarray} 
In the above equation, $f$ and $n$ are the fermionic and bosonic distribution 
functions, respectively.

\begin{figure}
\protect
\centerline{\psfig{figure=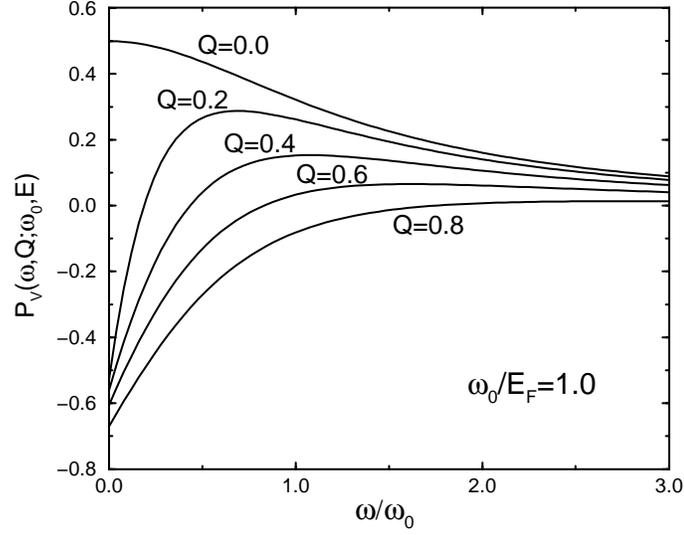,width=10cm}}
\caption{Vertex function $P$ as a function of the
exchanged frequency and for different values of the 
dimensionless momentum transfer $Q=q/(2k_F)$.} 
\label{fig2}
\end{figure}

By employing various approximations like for example the ones used
in Refs.\onlinecite{grima1,pietro3} (constant electron density of states,
structureless electron-phonon coupling and small momentum transfer) 
it is possible to perform analitycally the remaining summation
in Eq.(\ref{vertice1}). The result is shown in Fig. \ref{fig2}, where
we plot the vertex correction $P$ at half filling as a function of the exchanged
frequency $\omega_q$ and for different values of the dimensionless
momentum transfer $Q=q/(2k_F)$. For simplicity, the external
electron frequency $\omega_k$ has been set equal to zero.
In the figure we notice that $P$
can assume positive and negative values depending on the ratio
$v_F q/\omega_q$: for $v_F q/\omega_q>1$ the vertex is positive while
for $v_F q/\omega_q<1$ the vertex becomes negative.
This complex behavior is found also for more realistic band models\cite{perali} 
and it is also reflected in the different values 
the vertex assumes in the dynamic and static limits. In
fact, within the same approximation scheme used in the calculations
reported in Fig. \ref{fig2}, the static limit
$P_s=\lim_{\omega_q\rightarrow 0}P({\bf q}=0,\omega_q)$ is negative:
$P_s=-\omega_0/(\omega_0+E_F)$, while the dynamic one
$P_d=\lim_{{\bf q}\rightarrow 0}P({\bf q},\omega_q=0)$ is instead
positive: $P_d=E_F/(\omega_0+E_F)$.\cite{pietro2,grima1,pietro3}
The vertex correction is
therefore non-analytic in $\omega_q=0$, ${\bf q}=0$.

However this non-analyticity is removed when we consider the case  
of only one electron coupled to the lattice. In this situation
the electron distribution functions in (\ref{vertice1}) are striktly 
zero\cite{mahan}
and the vertex correction reduces to:

\begin{eqnarray}
\label{vertice2}
& & P({\bf k}+{\bf q},{\bf k};i\omega_k+i\omega_q,i\omega_k) =
\frac{\omega_0}{2}\sum_{{\bf k}'} g^2_{{\bf k}-{\bf k}'} \nonumber \\
&\times &\left[\frac{n(\omega_0)}{(\epsilon_{{\bf k}'}-\omega_0-i\omega_k)
(\epsilon_{{\bf k}'+{\bf q}}-\omega_0-i\omega_k-i\omega_q)}+
\frac{1+n(\omega_0)}{(\epsilon_{{\bf k}'}+\omega_0-i\omega_k)
(\epsilon_{{\bf k}'+{\bf q}}+\omega_0-i\omega_k-i\omega_q)}\right],
\end{eqnarray}
and it is straightforward to realize that the dynamic and static limits of
Eq.(\ref{vertice2}) are in fact equal. 

\begin{figure}
\protect
\centerline{\psfig{figure=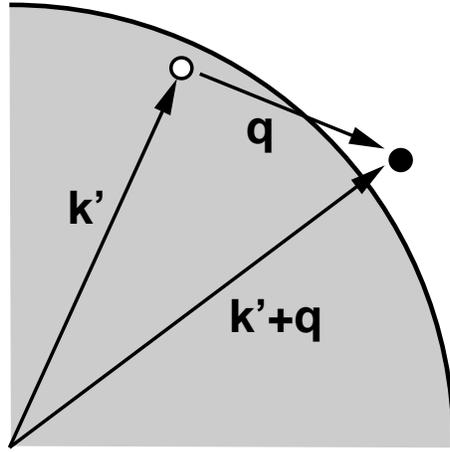,width=6cm}}
\caption{Representation of the particle-hole phase space.
The shaded area is the Fermi see, the open circle
an hole and the filled circle an electron.
For zero exchanged frequencies, particle-hole excitations 
can be obtained by connecting the hole with the electron
by the momentum transfer ${\bf q}$ such that 
${\bf q}\cdot{\bf k}'=0$.} 
\label{fig3}
\end{figure}

Therefore $P(k+q,k)$ has a strong dependence on the
exchanged frequency and momentum, and in particular is
non-analytic in ($\omega_q=0$,${\bf q}=0$), 
only when the electron density is finite. In fact, like in other
response functions, the non-analyticity of the vertex correction 
is given by hole-particle excitations which are 
present in Eq.(\ref{vertice1}).
These hole-particle processes can be explicitly singled out
by rearranging Eq.(\ref{vertice1}) in the following way:

\begin{eqnarray}
\label{vertice2bis}
& & P({\bf k}+{\bf q},{\bf k};i\omega_k+i\omega_q,i\omega_k) = 
P_{\mbox{pol}}({\bf k}+{\bf q},{\bf k};i\omega_k+i\omega_q,i\omega_k)+
P_{\mbox{ex}}({\bf k}+{\bf q},{\bf k};i\omega_k+i\omega_q,i\omega_k) 
\nonumber \\
&&=\frac{\omega_0}{2}\sum_{{\bf k}'}g^{2}_{{\bf k}-{\bf k}'}
\left[\frac{f(\epsilon_{{\bf k}'})+n(\omega_0)}
{(\epsilon_{{\bf k}'}-\omega_0-i\omega_k)
(\epsilon_{{\bf k}'+{\bf q}}-\omega_0-i(\omega_k+\omega_q))}+
\frac{1+n(\omega_0)-f(\epsilon_{{\bf k}'})}
{(\epsilon_{{\bf k}'}+\omega_0-i\omega_k)
(\epsilon_{{\bf k}'+{\bf q}}+\omega_0-i(\omega_k+\omega_q))}\right] \nonumber \\
&&-\sum_{{\bf k}'}g^{2}_{{\bf k}-{\bf k}'}
\frac{\omega_0^2}
{(i\omega_k+i\omega_q-\epsilon_{{\bf k}'+{\bf q}})^2-\omega_0^2} 
\frac{f(\epsilon_{{\bf k}'+{\bf q}})-f(\epsilon_{{\bf k}'})}
{\epsilon_{{\bf k}'+{\bf q}}-\epsilon_{{\bf k}'}-i\omega_q} .
\end{eqnarray}
The first term, $P_{\mbox{pol}}$, has equal dynamic and static 
limits and it reduces 
to Eq.(\ref{vertice2}) in the one electron case. In the next 
section we shall show that $P_{\mbox{pol}}$ stems from the
polarization of the lattice induced by the electrons and
it is basically a single electron process.
The second term of Eq.(\ref{vertice2bis}), $P_{\mbox{ex}}$, 
has instead a non-zero static limit and vanishes in the dynamic limit.
It is therefore this term that is responsible for the different
values of the dynamic and static limits of the vertex function 
(\ref{vertice1}) when the electron density is finite.
As it is clear from the expression in 
Eq.(\ref{vertice2bis}), the behavior of $P_{\mbox{ex}}$ 
is governed by particle-hole excitations since the factor

\begin{equation}
\label{p-h}
\frac{f(\epsilon_{{\bf k}'+{\bf q}})-f(\epsilon_{{\bf k}'})}
{\epsilon_{{\bf k}'+{\bf q}}-\epsilon_{{\bf k}'}-i\omega_q}=
\frac{f(\epsilon_{{\bf k}'+{\bf q}})[1-f(\epsilon_{{\bf k}'})]}
{\epsilon_{{\bf k}'+{\bf q}}-\epsilon_{{\bf k}'}-i\omega_q}+
\frac{f(\epsilon_{{\bf k}'})[1-f(\epsilon_{{\bf k}'+{\bf q}})]}
{\epsilon_{{\bf k}'}-\epsilon_{{\bf k}'+{\bf q}}+i\omega_q} ,
\end{equation}
describes particle-hole pairs creation.
The reason of having different values of the static and dynamic limits
is contained just in Eq.(\ref{p-h}). In fact, the particle-hole
excitation processes depend strongly on the available phase space
as it is shown in Fig. \ref{fig3} where we show schematically
the process given by the last term of Eq.(\ref{p-h}). For zero
exchanged frequency, $\omega_q=0$, particle-hole
excitations are present when 
$\epsilon_{{\bf k}'+{\bf q}}=\epsilon_{{\bf k}'}$, and this condition 
is obtained by placing the hole and the electron close to the Fermi
surface. On the other hand, when the exchanged frequency is nonzero,
the particle-hole excitations vanishes linearly with the momentum
transfer ${\bf q}$ when ${\bf q}\rightarrow 0$. 
In the following, we show that the physical process which leads
to particle-hole excitations in $P_{\mbox{ex}}$ is given by the
exchange part of the phonon mediated electron-electron interaction.

From the above discussion, we have seen that the two contributions
$P_{\mbox{pol}}$ and $P_{\mbox{ex}}$ of the vertex function
$P$ have different behaviors. Therefore the problem of finding the
physical interpretation of the vertex $P$ can be simplified by 
looking separately at $P_{\mbox{pol}}$ and $P_{\mbox{ex}}$.
This can be done by considering particular limiting cases.
For example, in the limit of one electron in the system,
the particle-hole contributions vanish so that $P_{\mbox{ex}}=0$
and we are left only with $P_{\mbox{pol}}$ 
in the form of Eq.(\ref{vertice2}).
On the other hand, as it is clear from Eq.(\ref{vertice2bis}),
when we employ $\omega_0=\infty$ limit 
(non-retarded phonon propagator) for a finite electron density,
the polarization part $P_{\mbox{pol}}$ vanishes and the vertex
is determined entirely by 
$\lim_{\omega_0\rightarrow\infty}P_{\mbox{ex}}$. 

The physical interpretation of the one electron and $\omega_0=\infty$
limits becomes straightforward if we introduce an external potential 
$U_{\mbox{ext}}$ which couples to the electron density.
In fact, when the coupling to the lattice is absent, 
this potential modifies the electron distribution in a way which
depends on the explicit form of $U_{\mbox{ext}}$. However, when the electrons
interact with the phonons, the response of the electrons to the
external potential changes because of the electron-phonon coupling. This
situation can be described in terms of an effective potential
$U_{\mbox{eff}}$. Actually, 
the vertex function is part of the effective potential,
so that we can interpret the vertex correction in terms of the
physically more intuitive $U_{\mbox{eff}}$.

In the next section therefore we study the response of the electrons
to the external potential $U_{\mbox{ext}}$ for both
the one electron case and the $\omega_0=\infty$ limit.

\section{limiting cases}
\label{limiting}

\subsection{One-electron case: $P_{\mbox{pol}}$}
\label{onele}

First we consider the case of only one electron in the system. 
In this situation, the particle-hole contributions of the
vertex function vanish and the second term of Eq.(\ref{vertice2bis}),
$P_{\mbox{ex}}$, is zero. Therefore the vertex is given by the
one electron limit of $P_{\mbox{pol}}$, Eq.(\ref{vertice2}), which
at zero temperature reduces to:

\begin{equation}
\label{vertice3}
P_{\mbox{pol}}({\bf k}+{\bf q},{\bf k};i\omega_k+i\omega_q,i\omega_k) =
\frac{\omega_0}{2}\sum_{{\bf k}'} \frac{g^2_{{\bf k}-{\bf k}'} }
{(\epsilon_{{\bf k}'}+\omega_0-i\omega_k)
(\epsilon_{{\bf k}'+{\bf q}}+\omega_0-i\omega_k-i\omega_q)}.
\end{equation}
As already pointed out before, the dynamic and static limits
of $P_{\mbox{pol}}$ coincide and, by employing the same approximation
scheme of Refs.\onlinecite{grima1,pietro3}, the ${\bf q}=0$,
$\omega_q=0$ limit becomes:

\begin{equation}
\label{verlim}
P_{\mbox{pol}}({\bf q}=0,\omega_q=0)=
\lambda\frac{E/2}{\omega_0+E},
\end{equation}
where we have neglected the external electron frequency 
$\omega_k$ and $E$ is the electronic bandwidth.

Our aim is to find the physical origin of Eq.(\ref{vertice3})
and to explain the reason why the limit in Eq.(\ref{verlim}) is positive.
To this end, we approach the problem 
by reasoning in terms of the electron response to an external
potential $U_{\mbox{ext}}$.\cite{nota1}

Let us for the moment neglect the electron-phonon interaction. 
For simplicity, we assume also that, in the absence of the
external perturbation $U_{\mbox{ext}}$, 
the electron wavefunction for the state ${\bf k}$ and energy
$\epsilon_{{\bf k}}$ is well
described by a simple plane-wave
$\psi_{{\bf k}}^0({\bf r})=\exp(i{\bf k}\cdot{\bf r})/\sqrt{V}$,
where $V$ is the volume.

\begin{figure}
\protect
\centerline{\psfig{figure=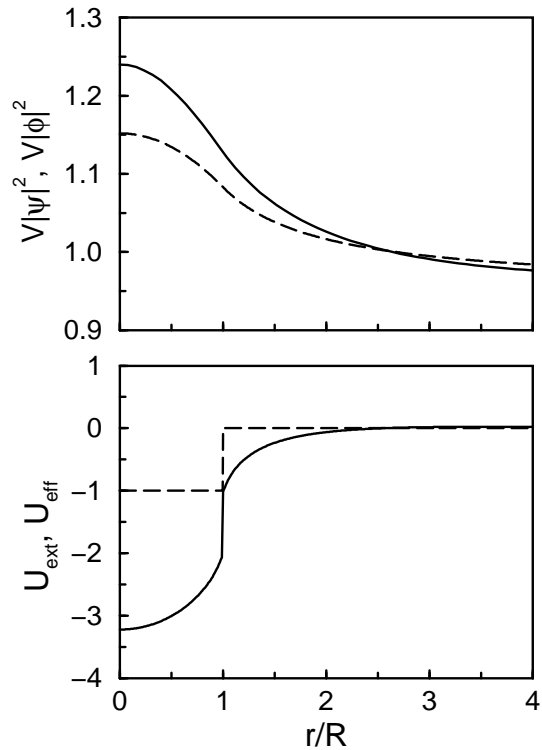,width=12cm}}
\caption{Upper panel: density of probability of 
one electron in the presence of an external potential
(plotted in the lower panel). Solid (dashed) lines: 
case with (without) electron-phonon interaction.
The potentials are plotted in units of $U_0$ and the
densities of probability are properly normalized.
In order to make clear the effect of the electron-phonon
coupling we have used suitable values of the
parameters entering in Eqs.(\protect\ref{wave2},\protect\ref{uffa2})
} 
\label{fig4}
\end{figure}

A nonzero external perturbation $U_{\mbox{ext}}$ modifies 
the electronic wavefunction
which, to the first order of the time-independent perturbation theory,
is given by:

\begin{equation}
\label{wave1}
\psi_{{\bf k}}({\bf r})=\psi_{{\bf k}}^0({\bf r})+\sum_{{\bf q}}
\frac{U_{\mbox{ext}}({\bf q})}{\epsilon_{{\bf k}}-\epsilon_{{\bf k}+{\bf q}}}
\psi_{{\bf k}+{\bf q}}^0({\bf r}).
\end{equation}
where $U_{\mbox{ext}}({\bf q})=\langle 
\psi_{{\bf k}+{\bf q}}^0|U_{\mbox{ext}}({\bf r})|\psi_{{\bf k}}^0\rangle$.
If we consider $U_{\mbox{ext}}({\bf r})$ to be given by a potential 
well of the form:

\begin{equation}
\label{ustat}
U_{\mbox{ext}}({\bf r})=\left\{ \begin{array}{cl}
-U_0   & r \leq R     \\
0 & r>R  \end{array}
\right. 
\end{equation}
then the density of probability $|\psi_{{\bf k}}({\bf r})|^2$
of finding a slow electron at position $r$ is enhanced inside the potential 
well and lowered outside the region $r\leq R$.
This is shown in
Fig. \ref{fig4} where we plot $V|\psi_{{\bf k}}({\bf r})|^2$
for ${\bf k}=0$ by a dashed line.
In the lower panel of the same figure 
we also plot the potential well $U_{\mbox{ext}}(r)$ of Eq.(\ref{ustat}) 
(dashed line).

Now we study how the above picture is modified when the electron
is weakly coupled to the lattice vibrations.
Intuitively, we expect that the
lattice is more polarized where the probability of finding
the electron is large, that is inside the potential well (for ${\bf k}$
small). 
The lattice polarization, in turns, provides an attractive potential
which is added to the external one. 
We can study this situation by replacing
$\psi_{{\bf k}}^0({\bf r})$ in Eq.(\ref{wave1}) by the wavefunction
modified by the coupling to the lattice vibrations:\cite{mahan}

\begin{equation}
\label{wave3}
\phi_{{\bf k}}^0({\bf r})=\psi_{{\bf k}}^0({\bf r})+
\sqrt{\frac{\omega_0}{2}}\sum_{{\bf k}'}
\frac{g_{{\bf k}-{\bf k}'}}{\epsilon_{{\bf k}}-\epsilon_{{\bf k}'}-\omega_0}
b^{\dagger}_{{\bf k}'-{\bf k}}\psi_{{\bf k}'}^0({\bf r}).
\end{equation}
In this way, the new electron wavefunction in the presence of
the external potential reduces to
                                         
\begin{equation}
\label{wave2}
\phi_{{\bf k}}({\bf r})=\phi_{{\bf k}}^0({\bf r})+\sum_{{\bf q}}
\frac{U_{\mbox{eff}}({\bf k}+{\bf q},{\bf k})}
{\epsilon({\bf k})-\epsilon({\bf k}+{\bf q})}
\phi_{{\bf k}+{\bf q}}^0({\bf r}),
\end{equation}
where $\epsilon({\bf k})$ is the electron dispersion modified by the
electron-phonon coupling and $U_{\mbox{eff}}({\bf k}+{\bf q},{\bf k})$
is the effective external potential which results from the lattice 
polarization:

\begin{equation}
\label{uffa2}
U_{\mbox{eff}}({\bf k}+{\bf q},{\bf k})=
\langle \phi_{{\bf k}+{\bf q}}^0|U_{\mbox{ext}}({\bf r})|
\phi_{{\bf k}}^0\rangle=
U_{\mbox{ext}}({\bf q})
\left[1+\frac{\omega_0}{2}\sum_{{\bf k}'}\frac{|g_{{\bf k}-{\bf k}'}|^2}
{(\epsilon_{{\bf k}'}-\epsilon_{{\bf k}}+\omega_0)
(\epsilon_{{\bf k}'+{\bf q}}-\epsilon_{{\bf k}+{\bf q}}+\omega_0)} \right].
\end{equation}
The second term in square brackets is just
the electron-phonon vertex correction (\ref{vertice3}) calculated for
$i\omega_k=\epsilon_{{\bf k}}$ and $i\omega_k+i\omega_q=
\epsilon_{{\bf k}+{\bf q}}$, {\it i.e.}, at the poles of the incoming and outcoming
electron lines of the vertex diagram of Fig. \ref{fig1}b.
The vertex correction for the one-electron case
is therefore part of the effective potential arising
from the lattice polarization. From the above discussion, the lattice
polarization should in general amplify the potential
seen by the electron.
This is confirmed by the numerical results reported in Fig.\ref{fig4},
where in the lower panel we plot the Fourier transform of the effective 
potential (\ref{uffa2}) for ${\bf k}=0$ (solid line). 
Moreover, the enhanced potential leads to an enhanced
probability of finding the electron in the vicinity of 
the potential well (solid line in the upper panel of Fig.\ref{uffa2}).

At this point it is straightforward to understand why the 
${\bf q}\rightarrow 0$ limit of the vertex function for one electron
is positive (see Eq.(\ref{verlim})). In fact, for ${\bf k}=0$, 
$\lim_{{\bf q}\rightarrow 0} P_{\mbox{pol}}({\bf q})=
\langle P_{\mbox{pol}}(r)\rangle$, where $\langle\cdots\rangle$ means
the average over the volume of the system, and

\begin{equation}
\label{average}
\langle P_{\mbox{pol}}(r)\rangle=
\frac{\langle U_{\mbox{eff}}(r)\rangle-\langle U_{\mbox{ext}}(r)\rangle}
{\langle U_{\mbox{ext}}(r)\rangle}>0.
\end{equation}

Of course the treatement of the problem followed in this section
does not consider the effect of the other electrons when the electron 
density $n$ is finite. This effect is partially contained in the general
expression of $P_{\mbox{pol}}$ given by Eq.(\ref{vertice2}) 
which in fact can be interpreted along the same 
lines followed in this section by taking into account the 
fermionic statistics of the electrons. 
However, $P_{\mbox{pol}}$ is basically a single electron process and
it belongs to the class of processes for which the same electron can
generate a phonon at a certain time $t$ and then absorb it at a
later time $t'$. Such kind of processes are consequence of the
retarded phonon propagation. 
On the other hand, for many electrons systems, a different kind of
processes are those for which an electron generates a phonon which
is absorbed by a different electron. This is a many-electrons 
process which is not contained in $P_{\mbox{pol}}$ but it is
given by $P_{\mbox{ex}}$, which is in fact determined by particle-hole
excitations.

In the next section we provide an interpretation for the physical
origin of $P_{\mbox{ex}}$. As anticipated before, we will introduce
an external potential and we will consider the resulting effective
potential by employing the $\omega_0\rightarrow\infty$ limit for
which $P_{\mbox{pol}}$ vanishes and the interpretation of
$P_{\mbox{ex}}$ is particularly simple.

\subsection{Anti-adiabatic limit: $P_{\mbox{ex}}$}
\label{roleofn}

Let us consider the electron-phonon coupled system in the limit
$\omega_0\rightarrow \infty$. In order to have non-trivial
results, we perform this limit in such a way that the quantity 
$g_{{\bf q}}^2$ remains finite. From Eq.(\ref{vertice2bis}),
$\lim_{\omega_0\rightarrow \infty}P_{\mbox{pol}}=0$ and the
vertex function reduces to:

\begin{equation}
\label{verticeinf}
\lim_{\omega_0\rightarrow \infty}P(k+q,k)=
\lim_{\omega_0\rightarrow \infty}
P_{\mbox{ex}}(k+q,k) =
\sum_{{\bf k}'}g^2_{{\bf k}-{\bf k}'}
\frac{f(\epsilon_{{\bf k}'})-f(\epsilon_{{\bf k}'+{\bf q}})}
{\epsilon_{{\bf k}'}-\epsilon_{{\bf k}'+{\bf q}}+i\omega_q}.
\end{equation}
As we have seen before, this term has different dynamic and static 
limits, and in particular we have:

\begin{eqnarray}
\label{dyna}
\lim_{\omega_q\rightarrow 0,{\bf q}\rightarrow 0}
P_{\mbox{ex}}(k+q,k) & = & 0,\\
\label{stat}
\lim_{{\bf q}\rightarrow 0,\omega_q\rightarrow 0}P_{\mbox{ex}}(k+q,k) & = &
\sum_{{\bf k}'}g^2_{{\bf k}-{\bf k}'}
\frac{d f(\epsilon_{{\bf k}'})}{d\epsilon_{{\bf k}'}}\simeq -\lambda ,
\end{eqnarray}
where the last equality holds true at small temperatures and
$\lambda$ is the electron-phonon coupling constant.
Although we can interpret the zero value of the dynamic limit as due to
the vanishing contribution of the hole-particle excitation contribution,
the reason why the static limit is negative remains unclear.
In this section we try to clarify this point by considering
the problem of the electron response to
an external potential which couples to the electron density.

The anti-adiabatic limit $\omega_0\rightarrow \infty$ 
transforms the electron-phonon interaction
into an effective non-retarded electron-electron interaction.
The Hamiltonian can be obtained by integrating out the
phononic degress of freedom and then taking the
$\omega_0\rightarrow \infty$ limit. The result is:

\begin{equation}
\label{haminf}
H=\sum_{{\bf k},\sigma}\epsilon_{{\bf k}}c^{\dagger}_{{\bf k}\sigma}
c_{{\bf k}\sigma}+\sum_{{\bf q}}U_{\mbox{ext}}({\bf q})n({\bf q})
-\frac{1}{2}\sum_{{\bf q}} g_{{\bf q}}^2 
n({\bf q})n(-{\bf q}),
\end{equation}
where $n({\bf q})$ is the electron density operator:

\begin{equation}
\label{enne1}
n({\bf q})=\sum_{{\bf k}\sigma}
c^{\dagger}_{{\bf k}+{\bf q}\sigma}c_{{\bf k}\sigma}.
\end{equation}
Now let us consider the response of the system to the
external potential $U_{\mbox{ext}}$. Since the electrons are interacting
through $g_{{\bf q}}^2$, the response depends in general on the
whole electron configuration. The simplest way in order to deal with this
problem is to perform the Hartree-Fock approximation in the four-operator
term in Eq.(\ref{haminf}). This approach is equivalent to consider 
anti-symmetrized states of $N$ independent one-electron wavefunctions
and leads to the random phase approximation for the effective potential
seen by the electrons.
When we apply the Hartree-Fock approximation, the interaction term in 
Eq.(\ref{haminf}) becomes:

\begin{equation}
\label{hfa}
-\frac{1}{2}\sum_{{\bf q}} g_{{\bf q}}^2 
n({\bf q})n(-{\bf q})\rightarrow
-\sum_{{\bf q}} g_{{\bf q}}^2 
\langle n(-{\bf q})\rangle n({\bf q})+
\sum_{{\bf q}}\sum_{{\bf k}\,{\bf k}'\,\sigma}
g^2_{{\bf q}}
\langle c^{\dagger}_{{\bf k}+{\bf q}\sigma}c_{{\bf k}'\sigma}\rangle
c^{\dagger}_{{\bf k}'-{\bf q}\sigma}c_{{\bf k}\sigma},
\end{equation}
and, after a manipulation of the momenta indexes, 
the Hamiltonian (\ref{haminf})
can be rewritten as follows:

\begin{equation}
\label{haminf2}
H=\sum_{{\bf k},\sigma}\epsilon_{{\bf k}}c^{\dagger}_{{\bf k}\sigma}
c_{{\bf k}\sigma}+
\sum_{{\bf q}}U_{\mbox{ext}}({\bf q})n({\bf q})+
\sum_{{\bf k}\,{\bf q}\sigma} V_{\mbox{HF}}({\bf k}+{\bf q},{\bf k}) 
c^{\dagger}_{{\bf k}+{\bf q}\sigma}c_{{\bf k}\sigma},
\end{equation}
where $V_{\mbox{HF}}$ is the Hartee-Fock potential in the presence
of the external perturbation:

\begin{equation}
\label{hfpot}
V_{\mbox{HF}}({\bf k}+{\bf q},{\bf k})=
-g_{{\bf q}}^2\langle n(-{\bf q})\rangle+
\sum_{{\bf k}'}g_{{\bf k}-{\bf k}'}^2
\langle c^{\dagger}_{{\bf k}'\sigma}c_{{\bf k}'+{\bf q}\sigma}\rangle .
\end{equation}
The first term in the right hand side is the Hartree contribution which
results from the potential generated by the electrons regardless the
specific electronic configuration. 
The last term of Eq.(\ref{hfpot}) is instead
the Fock contribution which treats the electrons as being dressed
by the exchange hole and it comes from the last term of Eq.(\ref{hfa}).
Below we show that the vertex function $P_{\mbox{ex}}$ originates from
the Fock contribution of Eq.(\ref{hfpot}), {\it i.e.}, from the
exchange term of the phonon mediated electron-electron interaction.

The effective potential given by the redistribution of the electrons
is readily obtained by the linear response theory. In fact, for
small values of $U_{\mbox{ext}}$, the
expectation values appearing in Eq.(\ref{hfpot}) can be rewritten
as:

\begin{eqnarray}
\label{exp1}
\langle n(-{\bf q})\rangle & = & \langle n(-{\bf q})\rangle_0+
\delta\langle n(-{\bf q})\rangle \\
\label{exp2}
\langle c^{\dagger}_{{\bf k}'\sigma}c_{{\bf k}'+{\bf q}\sigma}\rangle & = &
\langle c^{\dagger}_{{\bf k}'\sigma}c_{{\bf k}'+{\bf q}\sigma}\rangle_0 +
\delta\langle c^{\dagger}_{{\bf k}'\sigma}c_{{\bf k}'+{\bf q}\sigma}\rangle 
\end{eqnarray} 
where the first and the last terms in the right hand sides are the 
expectation values in the absence and in the
presence of the external potential, respectively.
From Eqs.(\ref{hfpot}-\ref{exp2}),
$V_{\mbox{HF}}$ can be rewritten as $V_{\mbox{HF}}^0+
\delta V_{\mbox{HF}}$ and the hamiltonian (\ref{haminf2}) becomes:

\begin{equation}
\label{haminf3}
H=H_{\mbox{HF}}^0+\sum_{{\bf q}\,{\bf k}\sigma}
U_{\mbox{eff}}({\bf k}+{\bf q},{\bf k})c^{\dagger}_{{\bf k}+{\bf q}\sigma}
c_{{\bf k}\sigma},
\end{equation}
where $H_{\mbox{HF}}^0$ is the hamiltonian in the Hartree-Fock
approximation for $U_{\mbox{ext}}=0$ and

\begin{equation}
\label{uff5}
U_{\mbox{eff}}({\bf k}+{\bf q},{\bf k})=
U_{\mbox{ext}}({\bf q})+\delta V_{\mbox{HF}}({\bf k}+{\bf q},{\bf k}).
\end{equation}
The above expression is a self-consistent equation 
 because $\delta V_{\mbox{HF}}$ depends
implicitly on  $U_{\mbox{eff}}$. 
We can repeat the calculations by assuming a time-dependent external
potential. In this case, by employing
the linear response theory applied to the
Hamiltonian (\ref{haminf3}), the self-consistent equation for
$U_{\mbox{eff}}$ is given by:

\begin{eqnarray}
\label{uffacon}
U_{\mbox{eff}}({\bf k}+{\bf q},{\bf k},i\omega_q) & = &
U_{\mbox{ext}}({\bf q},i\omega_q)-2g^2_{{\bf q}}
\sum_{{\bf k}'}\frac{f(\epsilon_{{\bf k}'})-f(\epsilon_{{\bf k}'+{\bf q}})}
{\epsilon_{{\bf k}'}-\epsilon_{{\bf k}'+{\bf q}}+i\omega_q}
U_{\mbox{eff}}({\bf k}'+{\bf q},{\bf k}',i\omega_q) \nonumber \\
& + & \sum_{{\bf k}'}g^2_{{\bf k}-{\bf k}'}
\frac{f(\epsilon_{{\bf k}'})-f(\epsilon_{{\bf k}'+{\bf q}})}
{\epsilon_{{\bf k}'}-\epsilon_{{\bf k}'+{\bf q}}+i\omega_q}
U_{\mbox{eff}}({\bf k}'+{\bf q},{\bf k}',i\omega_q),
\end{eqnarray}
where $\omega_q$ is the Matsubara frequency provided by the
time-dependence of the external potential and $\epsilon_{{\bf k}}$
is now the electron dispersion in the Hartree-Fock approximation.

Equation (\ref{uffacon}) is just the random phase 
approximation applied to the electron-phonon coupled system in the 
anti-adiabatic limit $\omega_0\rightarrow \infty$.
The second term of Eq.(\ref{uffacon}) represents the electron response
to $U_{\mbox{ext}}$ governed by the Hartree potential while the last
term is the correction due to the exchange potential. To the first order
in $U_{\mbox{ext}}$, the last term of Eq.(\ref{uffacon}) can be
rewritten as $P_{\mbox{ex}} U_{\mbox{ext}}$ where $P_{\mbox{ex}}$ is
the vertex correction in the antiadiabatic limit (\ref{verticeinf}).
Therefore $P_{\mbox{ex}}$ is just the Fock contribution to the
electronic response to the external potential when $\omega_0\rightarrow\infty$.

\begin{figure}
\protect
\centerline{\psfig{figure=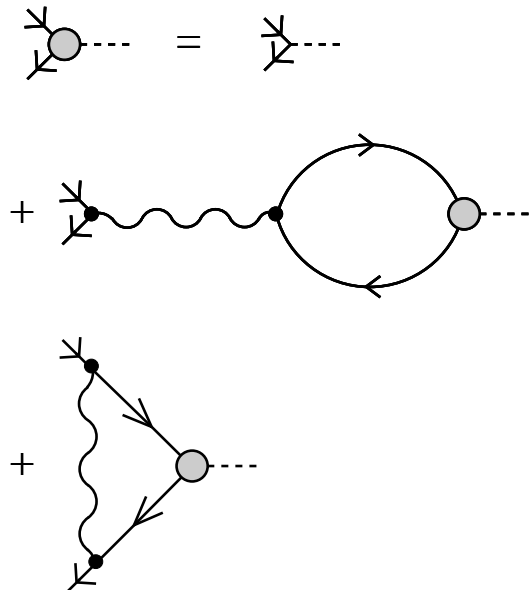,width=7cm}}
\caption{Diagramatic representation of the self-consistent
equation (\protect\ref{uffacon}).
The dashed circles represent the effective potential $U_{\mbox{eff}}$.
The second diagram comes from the Hartree interaction while
the last one from the Fock term. The wiggled lines represent
the phonon mediated electron-electron interaction in the antiadiabatic
limit.} 
\label{fig5}
\end{figure}

We show in Fig.\ref{fig5} the diagramatic respresentation of
the self-consistent equation (\ref{uffacon}). The Hartree term is
represented by the set of bubble diagrams while the Fock contribution
is given by the ladder contribution which, to the first order
in $U_{\mbox{ext}}$, is the vertex diagram.

At this point we can explain the behavior of the vertex function in
the antiadiabatic limit already
outlined in Eqs.(\ref{dyna},\ref{stat}). In fact,
the negative static limit of equation (\ref{stat}) 
can be understood by the following reasoning.
If we consider a static potential, then we must
set $\omega_q=0$ in Eq.(\ref{uffacon}). For simplicity let us also
consider an attractive potential like for example
the one in Eq.(\ref{ustat}).
If we neglect the Fock contribution, the electrons will tend to
form a cloud around the potential well and an added electron will
experience an effective potential given by the bare well plus the
electron cloud. Since the phonon mediated 
el-el interaction is attractive such an
effective potential will be stronger than the bare one.
However, when we consider the effect of Pauli principle, the
electron becomes dressed by the exchange hole which repels the
other electrons. Therefore the net effect of the Pauli
principle is to weaken the electron-electron attraction and
this weakening is reflected in the negative sign of the exchange term.
In conclusion, in the anti-adiabatic limit, the negative sign of
the vertex function for $\omega_q=0$, Eq.(\ref{stat}), is esclusively due to
the exchange effect which lower the phonon mediated
electron-electron interaction.

This picture is still valid when we take into account the retardation
of the phonon-mediated electron-electron interaction ($\omega_0<\infty$). 
In fact the last term of Eq.(\ref{vertice2bis}) can be interpreted as the 
Fock-like contribution for a retarded potential.

\section{Discussion and conclusions}
\label{material}

From the above analysis, we have seen that the vertex function
results from electron-phonon processes of different origins.
In fact, we have shown that the vertex $P$ can be decomposed
into two different contributions: $P_{\mbox{pol}}$ and $P_{\mbox{ex}}$.
The first one is the result of the lattice polarization as
induced by the electron motion. At low frequencies, 
this term leads to a positive contribution
to the electron-phonon interaction and tends to enhance the pairing.
Moreover, $P_{\mbox{pol}}$ is basically a single electron process.
The second term, $P_{\mbox{ex}}$, is instead due to the exchange
effect of the phonon mediated electron-electron interaction. Since
$P_{\mbox{ex}}$ is basically an exchange, it tends to reduce the
electron-phonon effective pairing. The presence of $P_{\mbox{ex}}$
is equivalent to have a lattice polarization effect due to an
electron which is moving with its exchange hole. Moreover, since
the exchange term gives rise to particle-hole excitations, 
$P_{\mbox{ex}}$ is sensitive to the momentum transfer and the 
exchanged frequency of the electron-phonon scattering process and it
gives rise to the different values of the dynamic and static limits of the
vertex function. In particular, in the dynamic limit the hole-particle
excitation vanish and the vertex is given only by its polarization part
$P_{\mbox{pol}}$, while in the static limit the particle-hole
excitations give rise to the exchange part which lower the effective
electron-phonon scattering.

From the above results and discussion, it is therefore straightforward to
identify scenarios in which the vertex correction can give rise to
an enhancement of 
the effective electron-phonon coupling. In fact, an enhancement can
be automatically obtained if the exchange effects become less
important.
This can be achieved, for example, when the charge carrier density
is low, so that the average distance between electrons can
exceed the size of the exchange hole leading to a negligible 
$P_{\mbox{ex}}$. In this way we can understand the results reported
in Ref.\onlinecite{free} where it is shown that, by reducing the
charge carrier density, the vertex correction
enhances the effective coupling. 

A more interesting situation in favour of an enhancement of the vertex corrected electron-phonon coupling is given by considering 
only small ${\bf q}$
scattering in the electron-phonon interaction. In fact,
for a small enough momentum transfer, say $v_F|{\bf q}|\ll\omega_0$, 
the particle-hole contributions (\ref{p-h}) have little weight
and the negative exchange term $P_{\mbox{ex}}$
becomes negligible. This result therefore clarifies
on physical grounds why the effective nonadiabatic electron-phonon
interaction and so the superconducting critical temperature $T_c$
is enhanced by the vertex corrections when the electron-phonon
interaction is only via small momentum transfer.\cite{grima1,pietro3}

It is important to stress that several works have shown how 
strong electronic correlations can lead to an electron-phonon 
interaction peaked at small momentum
transfer.\cite{szcz,kim,grilli,lee,zeyher}
Therefore strong electronic correlations can have a positive role
in enhancing the nonadiabatic effective electron-phonon
coupling by introducing a small upper cut-off on the momentum
transfer.
However, in the light of the results presented in this work,
we can more generally say that strong electronic
correlations tend to suppress the particle-hole excitations
consequently reducing the negative contribution of the exchange term
$P_{\mbox{ex}}$ in the vertex.

In conclusion, the two situations listed above, 
electron-phonon interaction with 
small momentum transfer (or more generally the suppression of the
particle-hole excitations due to strong electronic correlations)
and small carrier concentration, are both present in high-$T_c$ 
materials suggesting that the vertex corrections can enhance 
the pairing and give rise to high-$T_c$ superconductivity.

\acknowledgments
We would like to thank A. Amici, E. Cappelluti, S. Ciuchi and A. Perali
for useful discussions.
C. G. acknowledges the support of a INFM PRA project.

\end{document}